\documentclass[aps,prl,twocolumn,floatfix,showpacs]{revtex4}

\usepackage{bm,bbm,amsmath,times,amssymb,amsbsy,graphicx,amsthm,color,txfonts,pifont,natbib,hyperref}

\newcommand{\abl}[1]{{({#1})}}

 \newcommand{\tr}[1]{\text{Tr}[#1]}
\newcommand{\ket}[1]{|#1\rangle}
\newcommand{\bra}[1]{\langle#1|}
\newcommand{\proj}[1]{\ket{#1}\bra{#1}}

\newtheorem*{Theorem}{Theorem}

\begin{document}

\title
{Quantum Discord Determines the Interferometric Power of Quantum States}
% \\
% Quantum metrology embraced for the worst \\
% Employing quantum discord for phase information recovery}

% Place the author information here.  Please hand-code the contact
% information and notecalls; do *not* use \footnote commands.  Let the
% author contact information appear immediately below the author names
% as shown.  We would also prefer that you don't change the type-size
% settings shown here.

\author
{Davide Girolami$^{1,2,3}$, Alexandre M. Souza$^{4}$, Vittorio Giovannetti$^{5}$,
Tommaso Tufarelli$^{6}$, \\ Jefferson G. Filgueiras$^{7}$, Roberto S. Sarthour$^{4}$,
Diogo O. Soares-Pinto$^{8}$, Ivan S. Oliveira$^{4}$, Gerardo Adesso$^{1\ast}$}
\affiliation{%\quad \\
{$^{1}$School of Mathematical Sciences, The University of Nottingham,}
{University Park, Nottingham NG7 2RD, United Kingdom}\\
{$^{2}$Department of Electrical and Computer Engineering,
National University of Singapore, 4 Engineering Drive 3, Singapore 117583}\\
{$^{3}$Clarendon Laboratory, Department of Physics, University of Oxford, Parks Road, Oxford OX1 3PU, United Kingdom}\\
{$^{3}$Centro Brasileiro de Pesquisas F\'isicas, Rua Dr. Xavier Sigaud 150,}
{Rio de Janeiro, 22290-180  Rio de Janeiro, Brazil}\\
{$^{4}$NEST, Scuola Normale Superiore and Istituto Nanoscienze-CNR,}
{Piazza dei Cavalieri 7, I-56126 Pisa, Italy}\\
{$^{5}$QOLS, Blackett Laboratory, Imperial College London,}
{London SW7 2BW, United Kingdom}\\
{$^{6}$Fakult\"{a}t Physik, Technische Universit\"{a}t Dortmund, 44221 Dortmund, Germany} \\
{$^{7}$Instituto de F\'isica de S\~{a}o Carlos, Universidade de S\~{a}o Paulo,}
{P.O.~Box 369,  S\~{a}o Carlos, 13560-970 S\~{a}o Paulo, Brazil}\\
{$^\ast$To whom correspondence should be addressed; E-mail: \href{mailto:gerardo.adesso@nottingham.ac.uk}{gerardo.adesso@nottingham.ac.uk}}
}

% Include the date command, but leave its argument blank.

\date{May 16, 2014}

\pacs{03.65.Ud, 03.65.Wj, 03.67.Mn, 06.20.-f}

%%%%%%%%%%%%%%%%% END OF PREAMBLE %%%%%%%%%%%%%%%%


%\baselineskip24pt



\begin{abstract}
Quantum metrology exploits quantum mechanical laws to improve the precision in estimating technologically relevant parameters such as phase, frequency, or magnetic fields. Probe states are usually tailored to the particular dynamics whose parameters are being estimated. Here we consider a novel framework where quantum  estimation is performed in an interferometric configuration, using bipartite probe states prepared when only the spectrum of the generating Hamiltonian is known.  We introduce a figure of merit for the scheme, given by the worst-case precision over all suitable Hamiltonians, and prove that it amounts exactly to a computable measure of discord-type quantum correlations for the input probe. We complement our theoretical results with a metrology experiment, realized in a highly controllable room-temperature nuclear magnetic resonance setup, which provides a proof-of-concept demonstration for the usefulness of discord in sensing applications. Discordant probes are shown to guarantee a nonzero phase sensitivity for all the chosen generating Hamiltonians, while classically correlated probes are unable to accomplish the estimation in a worst-case setting.
This work establishes a rigorous and direct operational interpretation for general quantum correlations, shedding light on their potential for quantum technology.
\end{abstract}

\maketitle

%device independent
%min max

%Measurements are the fundamental tools
%for understanding and manipulating Nature.
 All quantitative sciences benefit from the spectacular developments in high accuracy  devices, such as atomic clocks, gravitational wave detectors and navigation sensors. Quantum metrology studies how to harness quantum mechanics to gain precision in estimating quantities not amenable to direct observation \cite{para,metrology,helstrom,susana,david}.  The phase estimation paradigm with measurement schemes based on an interferometric setup \cite{caves} encompasses a broad and relevant class of metrology problems, which can be conveniently cast in terms of an input-output
 scheme \cite{para}. An input probe state $\rho_{AB}$ enters a two-arm channel, in which the reference subsystem $B$ is unaffected while  subsystem $A$  undergoes a local unitary evolution, so that the output density matrix  can be written as $\rho_{AB}^{\varphi} = (U_A \otimes \mathbb{I}_B) \rho_{AB} (U_A \otimes \mathbb{I}_B)^{\dagger}$,
with %U^A=U_A\otimes\mathbb{I}_B,
 $U_A=e^{-i \varphi H_A}$,
  where $\varphi$ is the parameter we wish to estimate and $H_A$
   the local Hamiltonian generating the unitary dynamics.
 Information on~$\varphi$  is then recovered through  an estimator
function $\tilde{\varphi}$  constructed
 upon possibly joint measurements of suitable dependent observables performed on the output~$\rho_{AB}^\varphi$. % -- see Fig.~\ref{blinda}.
For any  input state $\rho_{AB}$ and generator $H_A$, the maximum achievable precision is determined theoretically by the  quantum Cram\'er-Rao bound~\cite{helstrom}. Given repetitive interrogations
via  $\nu$ identical copies of~$\rho_{AB}$,  this fundamental relation sets a lower limit to the mean square error  $\text{Var}_{\rho_{AB}^{\varphi}}(\tilde{\varphi})$ that measures the statistical distance between  $\tilde{\varphi}$ and $\varphi$: $\text{Var}_{\rho_{AB}^{\varphi}}(\tilde{\varphi})\geq \left[{\nu F(\rho_{AB};H_A)}\right]^{-1}$,
where $F$ is the quantum Fisher information (QFI) \cite{quantumcramer}, which quantifies how much information about $\varphi$ is encoded in $\rho_{AB}^\varphi$. The inequality is asymptotically tight as $\nu \rightarrow \infty$, provided the most informative quantum measurement is carried out at the output stage. Using  this quantity as a figure of merit, for independent and identically distributed trials, and under the assumption of complete prior knowledge of $H_A$, then {\it coherence} \cite{coherence} in the eigenbasis of $H_A$ is the essential resource for the estimation~\cite{metrology}; as maximal coherence in a known basis can be reached by a superposition state of subsystem $A$ only, there is no need for a correlated (e.g.~entangled) subsystem $B$ at all in this conventional case.

We show that the introduction of correlations is instead unavoidable when the assumption of full prior knowledge of $H_A$ is dropped. More precisely, we identify, in correlations commonly referred to as  {\it quantum discord} between $A$ and $B$ \cite{OZ,HV}, the necessary and sufficient resources rendering physical states able to store phase information in a unitary dynamics, independently of the specific Hamiltonian that generates it. Quantum discord is an indicator of quantumness of correlations in a composite system, usually revealed via the state disturbance induced by local measurements \cite{OZ,HV,activia,streltsov}; recent results suggested that discord might enable quantum advantages in specific computation or communication settings \cite{datta,npgu,npdakic,LQU,certif,pirliamo,modirev}. In this Letter a general quantitative equivalence between discord-type correlations and the guaranteed precision in quantum estimation is established theoretically, and observed experimentally in a liquid-state nuclear magnetic resonance (NMR) proof-of-concept implementation \cite{ernst,ivan}.

 %  \section*{Theory}

 %As we shall detail in the following, in this case  quantum correlations between $A$ and $B$ \cite{OZ,HV} become {\it the} fundamentalresource for estimation.   This can be well understood by considering
 %the following game of qubit systems, whose higher dimensional extension is presented in Methods \cite{epaps}.

\begin{figure}[t]
\begin{center}
\includegraphics[width=.37\textwidth]{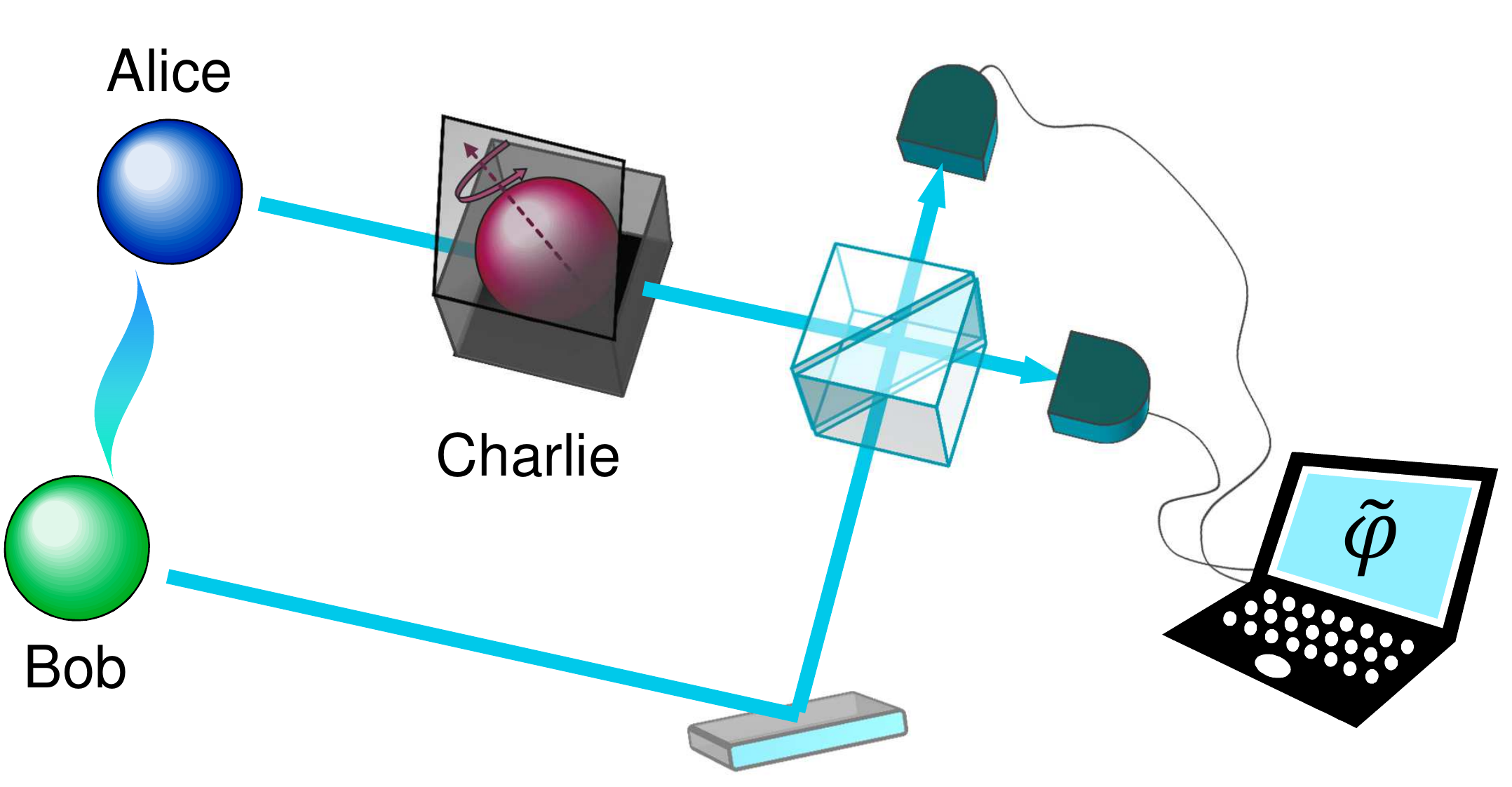}
\caption{(Color online) Black box quantum estimation.
\label{blinda}}
\end{center}
\end{figure}

 {\bf Theory.}
  An experimenter Alice, assisted by her partner Bob, has to determine as precisely as possible an unknown parameter $\varphi$ introduced by a {\it black box} device. The black box  implements the transformation $U_A=e^{-i \varphi H_A}$ and is controlled by a referee Charlie, see Fig.~\ref{blinda}. Initially, only the spectrum of the generator $H_A$ is publicly known and assumed to be nondegenerate. For instance, the experimenters might be asked to monitor a remote (uncooperative) target whose interaction with the probing signals is partially incognito \cite{QILLO}.
%We focus on the finite-dimensional case, although our analysis extends to continuous variable optical interferometry \cite{inprep}.
Alice and Bob prepare $\nu$ copies of a bipartite (generally mixed) probe state $\rho_{AB}$ of their choice. Charlie then distributes $\nu$ identical copies of the black box, and Alice sends each of her subsystems through one iteration of the box. After the transformations, Charlie reveals the Hamiltonian $H_A$ used in the box, prompting Alice and Bob to perform the best possible joint measurement on the transformed state $(\rho_{AB}^\varphi)^{\otimes \nu}$  in order to estimate $\varphi$ \cite{nmrnatural}.
%Alice and Bob prepare the systems $A$ and $B$ in a probe state $\rho_{AB}$. The probe enters the black box, where a randomizing mechanism, or an intelligent referee named Charlie, fixes a specific Hamiltonian $H_A$ on the spot and shifts subsystem $A$ by a phase $\varphi$.  Only after the transformation, Charlie can disclose the chosen Hamiltonian to Alice and Bob, who are prompted to perform the best possible joint measurement on the transformed state $\rho_{AB}^\varphi$  in order to estimate $\varphi$. The trial can be repeated an arbitrarily high number $\nu$ of times to improve the statistics, under the condition that the input probe state $\rho_{AB}$ and the chosen Hamiltonian $H_A$ are fixed by the first trial and not changed during the whole procedure.  This is a natural requirement for implementations such as NMR, where measurements on an ensemble of $\nu$ probes are taken collectively, rather than iteratively \cite{morton,nmrmetro,cory}.
Eventually, the experimenters infer a probability distribution associated to an optimal estimator $\tilde{\varphi}$ for $\varphi$ saturating the Cram\'er-Rao bound (for $\nu \gg 1$), so that the corresponding QFI determines exactly the estimation precision. For a given input probe state, a relevant figure of merit for this protocol is then given by the worst-case QFI over all possible black box settings,
\begin{equation}
\label{interpower}
{\mathcal P}^A(\rho_{AB}) = \frac{1}{4}\min_{H_A} F(\rho_{AB}; H_A)\,, %\vec n\cdot\vec \sigma_A)\,,
\end{equation}
where the minimum is intended over all Hamiltonians with given spectrum, and we inserted a normalization factor $\frac{1}{4}$ for convenience. We shall refer to ${\mathcal P}^A(\rho_{AB})$ as the {\it interferometric power} (IP) of the input state $\rho_{AB}$, since it naturally quantifies the guaranteed sensitivity that such a state allows in an interferometric configuration (Fig.~\ref{blinda}).

We prove that the quantity in Eq.~(\ref{interpower}) is a rigorous measure of discord-type quantum correlations of an arbitrary bipartite state $\rho_{AB}$   (see Supplementary Information \cite{epaps}).
If (and only if) the probe state is uncorrelated or only classically correlated, i.e.~Alice and Bob prepare a density matrix  $\rho_{AB}$ diagonal with respect to a local basis on $A$ \cite{activia,LQU,modirev}, then no precision in the estimation is guaranteed; indeed, in this case there is always a particularly adverse choice for $H_A$, such that $[\rho_{AB},H_A \otimes \mathbb{I}_B]=0$ and no information about $\varphi$ can be imprinted on the state, resulting in a vanishing IP. Conversely, the degree of discord-type correlations of the state $\rho_{AB}$ not only guarantees but also directly {\it quantifies}, via Eq.~(\ref{interpower}), its usefulness as a resource for  estimation of a parameter $\varphi$, regardless of the generator $H_A$ of a given spectral class.  For generic mixed probes $\rho_{AB}$, this is true even in absence of entanglement.

Remarkably, we can obtain a closed formula for the IP of an arbitrary quantum state of a bipartite system when subsystem $A$ is a qubit. Deferring the proof to \cite{epaps}, this reads
\begin{align}\label{IPfor}
	\mathcal P^{A}(\rho_{AB})=\varsigma_{\min}[M],
\end{align}
where $\varsigma_{\min}[M]$ is the smallest eigenvalue of the $3\times 3$ matrix $M$ of elements
\[
M_{m,n}=\frac{1}{2}\sum_{i,l:q_i+q_l\neq0}\frac{(q_i-q_l)^2}{q_i+q_l}\bra{\psi_i}{\sigma_m}_A\otimes\mathbb I_B\proj{\psi_l}{\sigma_n}_A\otimes\mathbb I_B\ket{\psi_i}\]
with $\{q_i, \ket{\psi_i}\}$ being respectively the eigenvalues and eigenvectors of $\rho_{AB}$, $\rho_{AB} = \sum_i q_i \ket{\psi_i}\!\bra{\psi_i}$.
This renders ${\mathcal P}^A(\rho_{AB})$ an operational {\it and} computable indicator of general nonclassical correlations for practical purposes.

\begin{figure}[t]
\begin{center}
%\vspace{0.1cm}
\includegraphics[width=8.5cm]{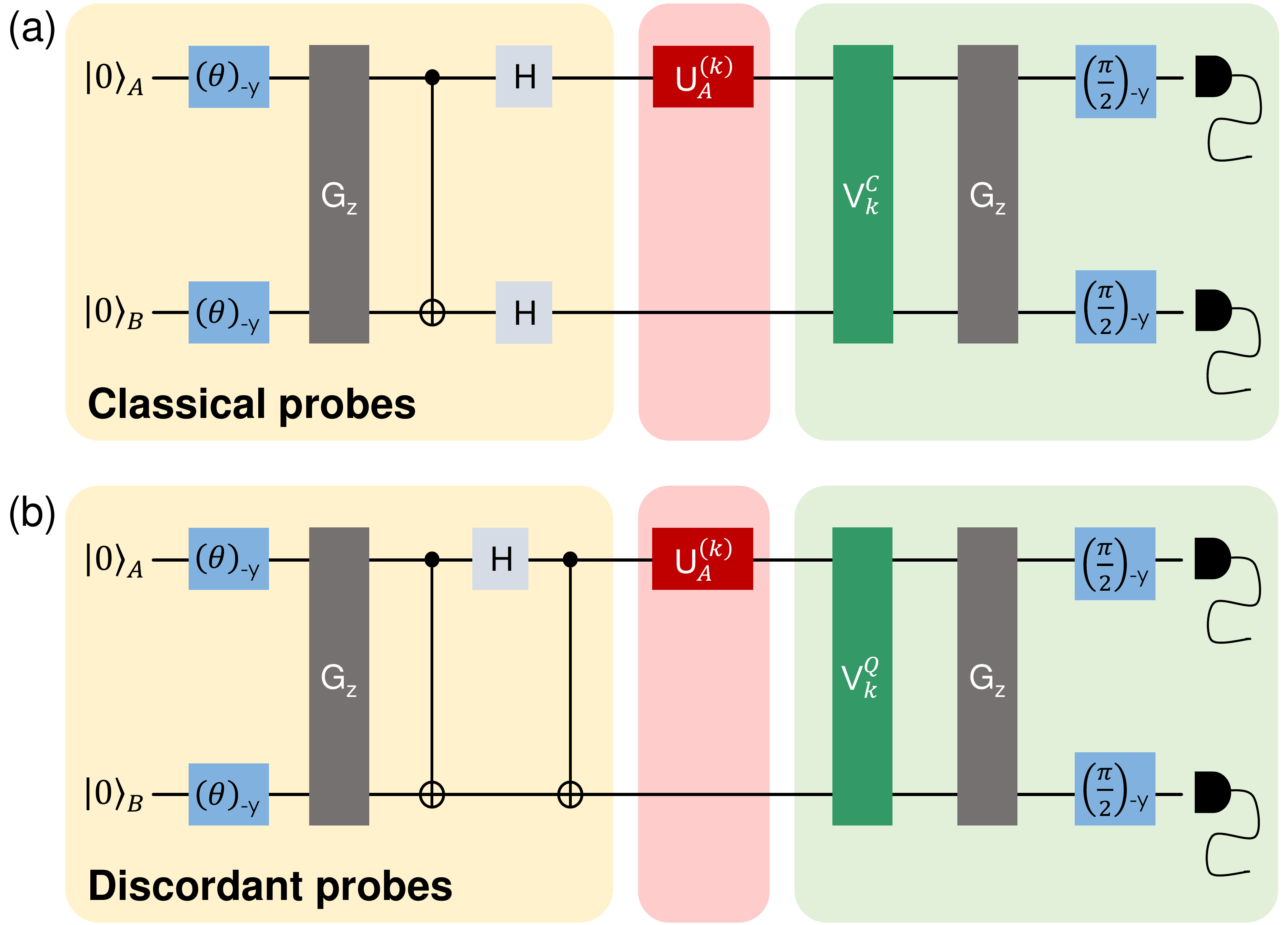}
\end{center}
\caption{(Color online) Experimental scheme for black box parameter estimation with NMR.
The protocol is divided in three steps: probe state preparation (yellow); black box transformation (red); optimal measurement (green).
%\abl{a} Graphical representation of the chloroform molecule. The $^1$H nuclear spin-$\frac12$ realises qubit $A$ and the $^{13}$C nuclear spin-$\frac12$ realises qubit $B$. \abl{b}
Starting from a thermal equilibrium distribution, we initialize the two-qubit system in a pseudo-pure state of the form $\rho =\frac{(1-\epsilon )}{4}\mathbb{I}+\epsilon~ \rho_{AB}$, with $\epsilon\sim 10^{-5}$ and  $\rho_{AB} = |00\rangle\langle 00|_{AB}$. This is done by applying the pulse sequence
$\big(\frac{\pi}{2}\big)_{x} \rightarrow U_{J}\big(\frac{1}{4J}\big) \rightarrow \big(\frac{\pi%
}{2}\big)_{y} \rightarrow U_{J}\big(\frac{1}{4J}\big) \rightarrow \big(\frac{\pi}{2}%
\big)_{-x} \rightarrow {\sf G}_{z} \rightarrow \big(\frac{\pi}{4}\big)_{y} \rightarrow U_{J}\big(\frac{1}{2J}\big) \rightarrow \big(\frac{\pi}{6}\big)_{x} \rightarrow {\sf G}_{z}$, where $\left(\theta\right)_{\alpha}$ is a rotation of each qubit by an
angle $\theta$ in the direction $\alpha$, $U_{J}\left(\tau\right)$ is a free
evolution under the scalar interaction between the spins for a time $\tau$, and
${\sf G}_{z}$ is a pulsed field gradient (which dephases  all the spins along the $z$ axis). We then proceed to prepare two types of probe states, the classically correlated ones $\rho^C_{AB}$ (a) and the discordant ones $\rho^Q_{AB}$ (b), defined in Eq.~(\ref{defstates}).
We first apply rf pulses, with a flip angle $\theta$, followed by a pulsed field gradient ${\sf G}_z$; this allows us to tune the purity parameter $p=\cos{\theta}$, by varying $\theta$ between $0$ and $90^\circ$ in steps of $2.5^\circ$. The subsequent circuits differ for each type of state: for $\rho_{AB}^C$ (a), a {\sf CNOT} gate followed by Hadamard gates $\sf H$ on both qubits $A$ and $B$ are implemented, while for $\rho_{AB}^Q$ (b), the {\sf CNOT} is followed  by a Hadamard $\sf H$ on qubit $A$ only and by a second {\sf CNOT}.
}
\label{figs2}
\end{figure}

\begin{figure*}[t]
\includegraphics[width=13cm]{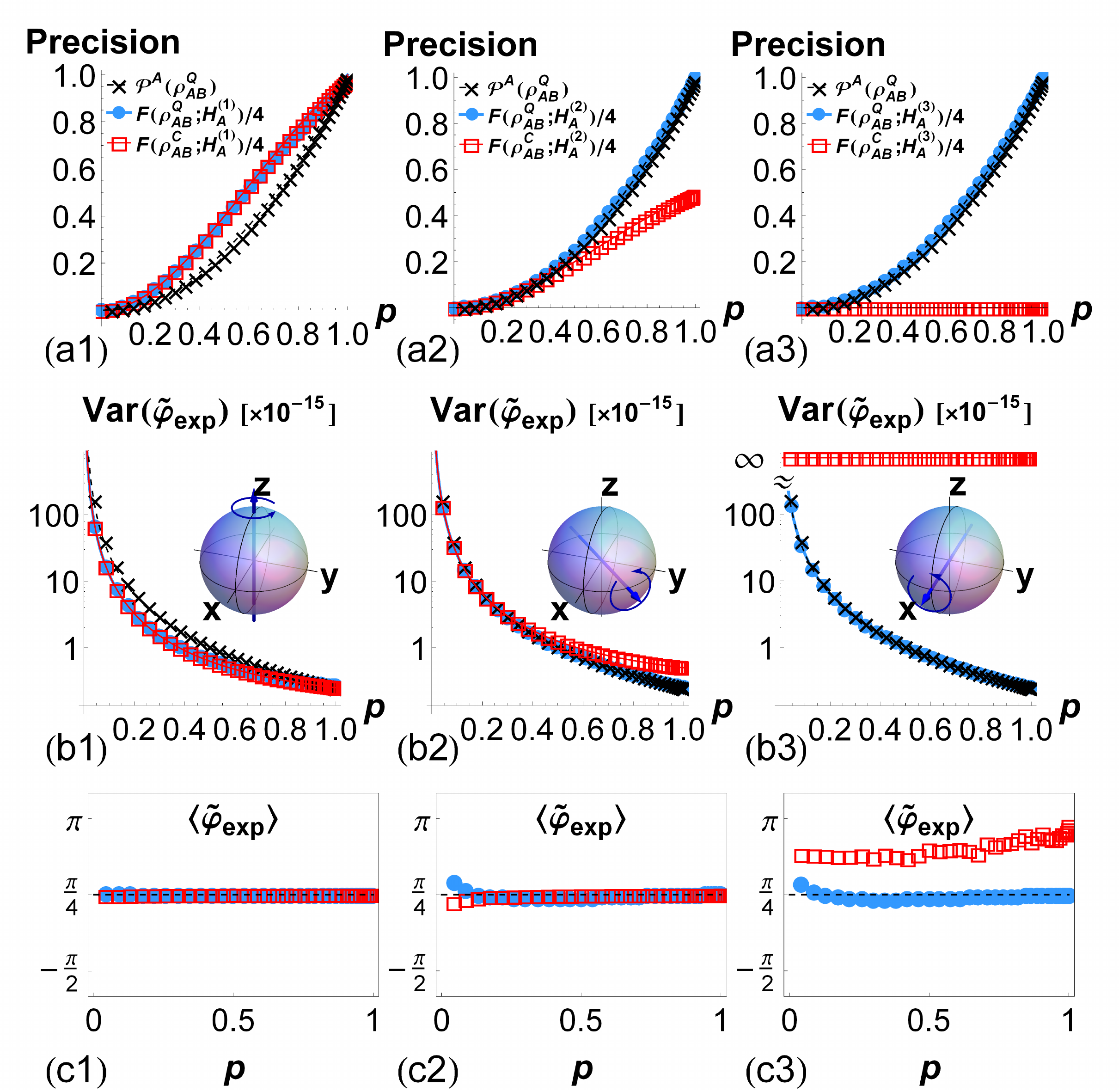}
\caption{(Color online) Experimental results. Each column corresponds to a different black box setting $H_A^{(k)}$, $k=1,2,3$, generating a $\varphi$ rotation on qubit $A$ around a Bloch sphere direction $\vec{n}^{(k)}$; the set directions are depicted in the insets of row (b). Empty red squares refer to data from classical probes $\rho_{AB}^C$,  filled blue circles refer to data from discordant probes $\rho_{AB}^Q$; error bars, due to small pulse imperfections in state preparation and tomography \cite{epaps}, are smaller than the size of the points. The lines refer to theoretical predictions.
Both families of states depend on a purity parameter $p$, experimentally tuned by a flip angle (see Fig.~\ref{figs2}).
The top row (a) shows the precision achieved by each probe in estimating $\varphi$ for the different settings: the  respective QFIs (divided by $4$) as obtained from the output measured data  are plotted and compared with the  IP ${\cal P}^A(\rho_{AB}^Q)$ of the discordant states (black crosses) measured from initial state tomography. The theoretical predictions are:
${F_{\rm th}(\rho^Q_{AB};H_A^{(1)})} = {F_{\rm th}(\rho^C_{AB};H_A^{(1)})} = \frac{8 p^2}{1+p^2}$, ${F_{\rm th}(\rho^Q_{AB};H_A^{(2)})} = 4 p^2,\, {F_{\rm th}(\rho^C_{AB};H_A^{(2)})} = \frac{4 p^2}{1+p^2}$, ${F_{\rm th}(\rho^Q_{AB};H_A^{(3)})} = 4 p^2,\, {F_{\rm th}(\rho^C_{AB};H_A^{(3)})} = 0$, and ${\cal P}^A(\rho^Q_{AB})=p^2$.
%Having no guaranteed lower bound (${\cal P}^A(\rho_{AB}^C)=0$), the classical probes may reach precisions as high as the quantum ones (as in case $k=1$) but can also fail the task completely (as in case $k=3$).
The middle row (b) depicts the measured variances of the optimal estimators $\tilde{\varphi}_{\rm exp}$ over the ensemble of $\nu \approx 10^{15}$ molecules, together with the theoretical predictions corresponding to the saturation of the quantum Cram\'er-Rao bound. The upper bound limiting the estimation uncertainty for the discordant states is shown as well, given by $4 [\nu {\cal P}^A(\rho_{AB}^Q)]^{-1}$ as calculated from the input states (crosses). The bottom row (c)  depicts the inferred mean value of the optimal estimator $\langle\tilde{\varphi}_{\rm exp}\rangle$ for the various settings.
 Both classical and quantum probes allow to get a consistently unbiased guess for $\varphi$
(in the experiment, the true value of $\varphi$ was set at $\varphi_0=\frac\pi 4$), apart from the unreliable results of $\rho_{AB}^C$ for $k=3$, which demonstrate that classical probes cannot return any estimation in the worst-case scenario.
%The pale red and blue shadings denote confidence intervals given by the (unrescaled) measured variances from row (b) for classical and discordant probes, respectively. The actual statistical uncertainties in our experiment amount to the plotted variances divided by the ensemble size $\nu \approx 10^{15}$.
%., which yields a remarkable precision even for small $p$.
}
\label{FigExp}
\end{figure*}

%\section*{Experiment}
{\bf Experiment.} We % complement our study by presenting
report an experimental implementation of black box estimation in a room temperature liquid-state NMR setting \cite{ivan,ernst}. Here quantum states are encoded in the spin configurations of magnetic nuclei of a $^{13}$C-labeled chloroform (CHCl$_3$) sample diluted in $d_6$ acetone. The  $^{1}$H and  $^{13}$C nuclear spins realize qubits $A$ and $B$, respectively; the states $\rho_{AB}$ are  engineerable as pseudo-pure states \cite{cory,nielsen} by controlling the deviation matrix from a fully thermal ensemble \cite{ivan}. A highly reliable implementation of unitary phase shifts can be obtained by means of radiofrequency (rf) pulses.
 Referring to \cite{epaps} for further details of the sample preparation and implementation, we now discuss the plan (Fig.~\ref{figs2}) and the results (Fig.~\ref{FigExp}) of the experiment.

%%%%%%%%%%%%%%%%%%%%%%%%%%%%%%%%%%%%%%%%%%%%%%%%%%%%%%%%%%%%%%%%%%%%%%%%%

  We compare two scenarios, where Alice and Bob prepare input probe states $\rho_{AB}$ either with or without discord. The chosen families of states are respectively \cite{nmrmetro,qstate,modix}
 \begin{equation}
 \begin{split}
 \rho_{AB}^Q&=\frac14\left(
\begin{array}{cccc}
1+p^2&0&0&2p\\
 0&1-p^2&0&0\\
0&0&1-p^2&0\\
 2p&0&0&1+p^2
\end{array}
\right), \\%\nonumber \\ & & \\
\rho_{AB}^C&=\frac14\left(
\begin{array}{cccc}
1& {p^2}& p&  p\\
 {p^2}&  1& p &  p \\
 p&  p&  1& {p^2}\\
 p& p&{p^2}& 1
\end{array}
\right).
\end{split} \label{defstates}
%\nonumber
\end{equation}
Both classes of states have the same purity, given by $\text{Tr}\Big[(\rho_{AB}^{Q,C})^2\Big] = \frac14(1+p^2)^2$, where $0\leq p \leq 1$. This allows us to focus on the role of initial correlations for the subsequent estimation, at tunable common degree of mixedness mimicking realistic environmental conditions. While the states $\rho_{AB}^{C}$ are classically correlated for all values of $p$, the states $\rho_{AB}^{Q}$ have  discord increasing monotonically with $p>0$. The probes are prepared by applying a  chain of control operations to the initial Gibbs state (Fig.~\ref{figs2}).
%, and are furthermore entangled for $p>\sqrt{2}-1$.
We perform a full tomographical reconstruction of each input state to validate the quality of our state preparation, obtaining a mean fidelity of $(99.7 \pm 0.2)\%$ with the theoretical density matrices of Eq.~(\ref{defstates}) \cite{epaps}. In the case of $\rho_{AB}^{Q}$, we measure the degree of discord-type correlations in the probes by evaluating the closed formula (\ref{IPfor}) for the IP  on the tomographically reconstructed input density matrices; this is displayed as black crosses in the top panel of Fig.~\ref{FigExp} and is found in excellent agreement with the theoretical expectation ${\cal P}^A(\rho_{AB}^Q)=p^2$.

  Then, for each fixed input probe, and denoting by $\varphi_0$ the true value of the unknown parameter $\varphi$ to be estimated by Alice and Bob (which we set to $\varphi_0=\frac{\pi}{4}$ in the experiments without any loss of generality),  we implement three different choices of Charlie's black box transformation $U_A^{(k)}=e^{-i \varphi_0 H_{A}^{(k)}}\otimes \mathbb{I}_B$. These are given by $H_A^{(1)}={\sigma_z}_A$, $H_A^{(2)}=({\sigma_x}_A+{\sigma_y}_A)/\sqrt2$, $H_A^{(3)}={\sigma_x}_A$, and are respectively engineered by applying the pulse sequences $U_A^{(1)}=\big(\frac{\pi}{2}\big)_{x} \rightarrow \big(\frac{\pi}{2}\big)_{-y} \rightarrow \big(\frac{\pi}{2}\big)_{-x}, U_A^{(2)}=\big(\frac\pi 2\big)_{x+y}, U_A^{(3)}= \big( \frac{\pi }{2}\big) _{x}$.  A theoretical analysis asserts that the chosen settings encompass the best (setting $k=1$) and worst (setting $k=3$) case scenarios for both types of probes, while the setting $k=2$ is an intermediate case \cite{epaps}.

For each input state and black box setting, we carry out the corresponding optimal measurement strategy for the estimation of $\varphi$.
This is given by projections on the eigenbasis $\{\ket{\lambda_j}\}$ ($j=1,\ldots,4$) of the symmetric logarithmic derivative (SLD) $L_{\varphi}=\sum_j l_j \ket{\lambda_j}\bra{\lambda_j}$, an operator satisfying $\partial_{\varphi}\rho_{AB}^{\varphi}=\frac{1}{2} (\rho_{AB}^{\varphi}L_{\varphi}+L_{\varphi}\rho_{AB}^{\varphi})$ \cite{quantumcramer}. The QFI is then given by \cite{paris} $F(\rho_{AB};H_A)=\tr{\rho_{AB}^{\varphi}L_{\varphi}^2}= 4 \sum_{i<l:q_i+q_l\neq0} \frac{(q_i-q_l)^2}{q_i + q_l} |\bra{\psi_i}(H_A \otimes \mathbb{I}_B)\ket{\psi_l}|^2$, where $\{q_i, \ket{\psi_i}\}$ are the eigenvalues and eigenvectors of $\rho_{AB}$ as before. We implement a readout procedure based on a global rotation into the eigenbasis of the SLD, depicted as $V_k^{C,Q}$ in Fig.~\ref{figs2}, followed by a pulsed field gradient ${\sf G}_z$ to perform an ensemble measurement of the expectation values $d_j = \bra{\lambda_j} \rho_{AB}^\varphi \ket{\lambda_j}$ averaged over $\nu \approx 10^{15}$ effectively independent probes \cite{nmrmetro,epaps}. These are read from the main diagonal of the output density matrices, circumventing the need for complete state tomography. The measurement basis is selected by a simulated adaptive procedure and the measured ensemble data $d_j^{\rm exp}$ are reported in \cite{epaps}.

To accomplish the estimation, we need  a statistical estimator for $\varphi$. Denoting by $(k,s)$ an instance of the experiment (with $k=1,2,3$ referring to the black box setting, and $s=C,Q$ referring to the input probes), an optimal estimator for  $\varphi$ which asymptotically saturates the quantum Cram\'er-Rao bound can be formally constructed as \cite{paris,notaav}
\begin{equation}\label{imbroglio}
\tilde{\varphi}^{(k,s)} = \varphi_0 \mathbb{I} + \frac{L^{(k,s)}_{\varphi_0}}{\sqrt{\nu} F(\rho^s_{AB};H_A^{(k)})}\,,
\end{equation}
such that
$\langle \tilde{\varphi}^{(k,s)} \rangle = \varphi_0$, and $\text{Var}(\tilde{\varphi}^{(k,s)}) = [{\nu F(\rho^s_{AB};H_A^{(k)})}]^{-1}$, because $\langle L^{(k,s)}_{\varphi_0} \rangle = 0$ by definition. However, the estimator in Eq.~(\ref{imbroglio}) requires the knowledge of the true value $\varphi_0$ of the unknown parameter, which cannot be obtained by iterative procedures in our setup. We then infer directly the ensemble mean  and variance of the optimal estimator from the available data, namely the measured values $d_j^{{\rm exp}(k,s)}$, and the knowledge of the input probe states $\rho_{AB}^s$ prepared (and reconstructed) by Alice and Bob, of the setting $k$ disclosed by Charlie, and of the design eigenvalues $l_j$ of the SLD, which are independent of $\varphi$.

First, we infer the expected value of the optimal  estimator $\tilde{\varphi}^{(k,s)}$ by means of a statistical least-squares processing. We derive theoretical model expressions for the measured data $d_j^{{\rm exp}(k,s)}$ defined by
$d_j^{{\rm {th}}(k,s)}(\varphi) = \bra{\lambda_j^{\varphi_0 (k,s)}} (e^{-i \varphi H_A^{(k)}} \otimes \mathbb{I}_B) \rho^s_{AB} (e^{i \varphi H_A^{(k)}} \otimes \mathbb{I}_B) \ket{\lambda_j^{\varphi_0 (k,s)}}$,
and calculate the value of $\varphi$ which minimizes the least-squares function
$\Upsilon^{(k,s)}(\varphi)=\sum_{j=1}^{4} \left[d_j^{{\rm {th}}(k,s)}(\varphi)-d_j^{{\rm {exp}}(k,s)}\right]^2$ (equivalent to maximizing the log-likelihood assuming that each $d_j$ is Gaussian-distributed over the ensemble). For each setting $(k,s)$, the value of $\varphi$ that solves the least-squares problem is chosen as the expected value $\langle \tilde{\varphi}_{\rm exp}^{(k,s)} \rangle$ of our estimator. These values are plotted in row (c) of Fig.~\ref{FigExp}: one can appreciate the agreement with the true value $\varphi_0=\frac{\pi}{4}$ of the unknown phase shift for all settings but the pathological one $(3,C)$. In the latter case, the estimation is completely unreliable because the classical probes commute with the corresponding Hamiltonian generator, thus failing the  estimation task.

Next, by expanding the SLD in its eigenbasis (see Table~S--I in \cite{epaps}), we obtain the experimental QFIs measured from our data,
${F_{\rm exp}(\rho^s_{AB};H_A^{(k)})} = \langle \big({L^{(k,s)}_{\varphi_0}}\big)^2 \rangle=
\sum_j (l_j^{(k,s)})^2\ d_j^{{\rm {exp}}(k,s)}$.
These are plotted (normalized by a factor $\frac14$) for the various settings in row (a) of Fig.~\ref{FigExp}, together with the lower bound given by the IP of $\rho_{AB}^Q$. We remark that the QFIs are obtained from the output estimation data, while the IP is measured on the input probe states. In both cases an excellent agreement with theoretical expectations is retrieved for all settings. Notice how in cases $k=2,3$ the quantum probes achieve a QFI that saturates the lower bound given by the IP. Notice also that for $k=3$ the classical states yield strictly zero  QFI, as $l_j^{(3,C)}=0$ $\forall j$.

 %inputexisting for  %of Figure 3 of the main text,
%again in accordance with the theoretical predictions, which are given explicitly by
%\begin{eqnarray}\label{tfish}
%{F_{\rm th}(\rho^Q_{AB};H_A^{(1)})} = \frac{8 p^2}{1+p^2}, &\qquad& {F_{\rm th}(\rho^C_{AB};H_A^{(1)})} = \frac{8 p^2}{1+p^2}, \nonumber \\
%{F_{\rm th}(\rho^Q_{AB};H_A^{(2)})} = 4 p^2, &\qquad& {F_{\rm th}(\rho^C_{AB};H_A^{(2)})} = \frac{4 p^2}{1+p^2}, \\
%{F_{\rm th}(\rho^Q_{AB};H_A^{(3)})} = 4 p^2, &\qquad& {F_{\rm th}(\rho^C_{AB};H_A^{(3)})} = 0.
%\end{eqnarray}

Finally, we infer the variance of the optimal estimator  over the spin ensemble. This is obtained by replacing $\varphi_0 \mathbb{I}$ with $\langle \tilde{\varphi}_{\rm exp}^{(k,s)} \rangle \mathbb{I}$ in Eq.~(\ref{imbroglio}) and calculating $\text{Var}(\tilde{\varphi}_{\rm exp}^{(k,s)})$ by expanding it in terms of the design weight values $l_j^{(k,s)}$ and the measured data $d_j^{{\rm {exp}}(k,s)}$; namely, $\text{Var}(\tilde{\varphi}_{\rm exp}^{(k,s)}) = \big[\big(\sum_j (l_j^{(k,s)})^2\ d_j^{{\rm {exp}}(k,s)}\big) - \big(\sum_j l_j^{(k,s)}\ d_j^{{\rm {exp}}(k,s)}\big)^2\big]/\big[\nu \big({F_{\rm exp}(\rho^s_{AB};H_A^{(k)})}\big)^2\big]$. The resulting variances $\text{Var}(\tilde{\varphi}_{\rm exp}^{(k,s)})$  of our  metrology experiment are then plotted in  row (b) % and as confidence intervals in row  (c)
of Fig.~\ref{FigExp}.
%(without the statistical rescaling factor $\nu$ for ease of display).
The obtained quantities are in neat agreement with the inverse relation $\text{Var}(\tilde{\varphi}_{\rm exp}^{(k,s)}) \approx [{\nu F_{\rm exp}(\rho^s_{AB};H_A^{(k)})}]^{-1}$, which allows us to conclude that the implemented estimator with experimentally determined mean $\langle\tilde{\varphi}_{\rm exp}^{(k,s)}\rangle$ and variance $\text{Var}(\tilde{\varphi}_{\rm exp}^{(k,s)})$, constructed from our ensemble data, saturates the quantum Cram\'er-Rao bound: this confirms that an optimal detection strategy was carried out in all settings.
Overall, this clearly shows that discord-type quantum correlations, which establish {\it a priori} the guaranteed precision for any bipartite probe state via the quantifier ${\cal P}^A$, are the key resource for black box estimation, demonstrating the central claim of this Letter.

%   \section*{Conclusion}

{\bf Conclusion.} In summary, we investigated black box parameter estimation as a metrology primitive. We introduced the IP of a bipartite quantum state, which measures its ability to store phase information in a worst-case scenario. This was proven equivalent to a measure of the general quantum correlations of the state. We demonstrated the operational significance of discord-type correlations by implementing a proof-of-concept NMR black box estimation experiment, where the high controllability on state preparation and gate implementation allowed us to retain the hypothesis of unitary dynamics, and to verify the saturation of the Cram\'er-Rao bound for optimal estimation.
%Our study provides a quantitative framework to interpret discord as coherence \cite{coherence,activia,LQU} in all local bases. For every application relying on coherence in a certain basis, a bipartite strategy involving discordant states would ensure success even without knowing the preferred basis {\it a priori}. Phase estimation, as considered in this work, is one such task, where the quality of the process was linked to discord in a fully quantitative fashion. Other protocols where discord can be anticipated to play an essential prerequisite role are, e.g.,  channel discrimination \cite{inprep}, data hiding, and quantum key distribution \cite{pirla}.
%Our study goes in the direction of device-independent quantum information processing \cite{devind}, which for metrology
%can be fulfilled by reducing the assumptions on the manufacturers of the black box to be estimated. Here
%we achieved the first step of the programme, by lifting the initial knowledge on the specifics of the generator.
%Attempting to estimate a dynamical parameter without being disclosed its generator even after the transformation would realise a fully blind metrology setting %\cite{blind}, and require techniques of multiparametric estimation.  From the practical point of view,
Our results suggest that in highly disordered settings, e.g.~NMR systems, and under adverse conditions, quantum correlations even without entanglement can be a promising resource for realizing quantum technology.

%\bigskip

%\bigskip

%\section*{ }
\section*{Acknowledgments}
We acknowledge discussions with S.~Benjamin, T.~Bonagamba, T.~Bromley, M.~Cianciaruso, L.~Correa, L.~Davidovich, E.~deAzevedo, B.~Escher, E.~Gauger, M.~Genoni, M.~Guta, S.~Huelga, M.~S.~Kim, M.~Lang, B.~Lovett, K.~Macieszak, K.~Modi, J.~Morton, M.~Paris, M.~Piani, R.~Serra, M.~Tsang.
This work was supported by the Singapore National Research Foundation under NRF Grant No. NRF-NRFF2011-07,  the Foundational Questions Institute (FQXI), the University of Nottingham [EPSRC Research Development Fund Grant No.~PP-0313/36], the Italian Ministry of University and Research [FIRB-IDEAS Grant No.~RBID08B3FM], the Qatar National Research Fund [NPRP 4-426 554-1-084], the Brazilian funding agencies CAPES [Pesquisador Visitante Especial-Grant No.~108/2012], CNPq [PDE Grant No.~236749/2012-9], FAPERJ, and the Brazilian National Institute of Science and Technology of Quantum Information (INCT/IQ).

%\bibliographystyle{Science}

%\end{document}
%TODO EPAPS

\clearpage

\setcounter{page}{1}
\begin{widetext}
\medskip
%\appendix*

\section*{Supplementary Information}

\begin{center}
{\large{\bf {Quantum discord determines the interferometric power of quantum states}}}

\quad \\

{\normalsize Davide Girolami, Alexandre M. Souza, Vittorio Giovannetti,
Tommaso Tufarelli, \\ Jefferson G. Filgueiras, Roberto S. Sarthour,
Diogo O. Soares-Pinto, Ivan S. Oliveira, Gerardo Adesso}

\end{center}

\setcounter{equation}{0}

%% HACKS %%

% For section headers starting with S
%\renewcommand{\thesection}{S.\arabic{section}}
%\renewcommand{\thesubsection}{\thesection.\arabic{subsection}}

% Hack for making SOM Equations Conform to Science Format
%
% e.g. (S1), (S2), etc
% Requires AMS
%\makeatletter %% With ams
%\def\tagform@#1{\maketag@@@{(S\ignorespaces#1\unskip\@@italiccorr)}}
%\makeatother

\renewcommand{\theequation}{S\arabic{equation}}

% Hack for making figures Say \figurename S\thefigure, e.g. Figure S1:
%\makeatletter
%\makeatletter \renewcommand{\fnum@figure}
%{\figurename~S\thefigure}
%\makeatother

%\makeatletter
%\renewcommand{\thetable}{S-\@Roman\c@table}

% use bibnumfmt to change style at the end of the document
\renewcommand{\bibnumfmt}[1]{[S#1]}
% citenumfont command adds S to all numbers
\renewcommand{\citenumfont}[1]{{S#1}}

\renewcommand{\figurename}{Supplementary Figure}

\renewcommand{\tablename}{Supplementary Table}

\setcounter{figure}{0}
\setcounter{equation}{0}
\setcounter{table}{0}

\makeatletter
\renewcommand{\thefigure}{S\@arabic\c@figure}
\makeatother

\makeatletter
\renewcommand{\thetable}{S-\@Roman\c@table}
%END HACKS

\section{Theoretical properties of the interferometric power}
\subsection{Definition and evaluation of $\boldsymbol{{\cal P}^A}$}

We define the IP, in general, as
$$
	{\mathcal P}^A(\rho_{AB}|\Gamma) = \frac{1}{4}\min_{H_A^\Gamma} F(\rho_{AB}; H_A^\Gamma),
$$
where $F$ denotes the QFI
$F(\rho_{AB};H_A^\Gamma)= 4 \sum_{i<k:q_i+q_k\neq0} \frac{(q_i-q_k)^2}{q_i + q_k} |\bra{\psi_i}(H_A^\Gamma \otimes \mathbb{I}_B)\ket{\psi_k}|^2$, with $\{q_i, \ket{\psi_i}\}$ being respectively the eigenvalues and eigenvectors of $\rho_{AB}$, $\rho_{AB} = \sum_i q_i \ket{\psi_i}\!\bra{\psi_i}$,
and  the minimum is taken over the set of all local Hamiltonians $H_A^\Gamma$ with fixed nondegenerate spectrum $\Gamma=\{\gamma_i\}$. The mentioned set of Hamiltonians can be parameterized as $H_A^{\Gamma}=V_A\Gamma V_A^{\dagger}$, where $\Gamma=\text{diag}(\gamma_1,\ldots, \gamma_{d_A})$ is a fixed  diagonal matrix with ordered (non-decreasing) eigenvalues, while  $V_A$ can vary arbitrarily over the special unitary group SU$(d_A)$. In general, we expect different choices of $\Gamma$ to lead to different measures which induce  inequivalent orderings on the set of quantum states. Yet using the fact that, for all real constants $a$ and $b$, one has $F(\rho_{AB}; a H^{\Gamma}_A+b \mathbb I_A) = a^2 F(\rho_{AB}; H^{\Gamma}_A)$, one can transform $\Gamma$ to a canonical form where $\gamma_1=\frac{d_A}{2}$  and $\gamma_{d_A} =-\frac{d_A}{2}$.
%\textcolor{red}{The assumption of non-degeneracy of the Hamiltonian filters out the  ignorance on its spectrum, which we deem as lack of knowledge due to our inability to deal with  technical imperfections, and thus as {\it classical} ignorance, and take in account only the lack of knowledge on the eigenbasis of $H_A$, i.e.  the {\it quantum} uncertainty on it \cite{LQU}. This is necessary for addressing the resource which allows to recover phase information, i.e. information on the eigenbasis}. % (or other convenient choices).

Focusing now on the relevant case of subsystem  $A$ being a qubit, with $B$ a quantum system of arbitrary dimension, the only nontrivial canonical form for $\Gamma$ is $\text{diag}(1,-1)$, which corresponds to the set (we drop the superscript $\Gamma$ from now on)  $H_A=\vec n\cdot\vec{\sigma}_A$ with $|\vec n|=1$ and $\vec{\sigma}_A = ({\sigma_x}_A, {\sigma_y}_A, {\sigma_z}_A)$ being the vector of Pauli matrices. One can reduce the expression of ${\cal P}^A(\rho_{AB})$ to the minimization of a quadratic form over the unit sphere, hence obtaining the analytic expression announced in the main text:
\begin{align}\label{IPforS}
	\mathcal P^{A}(\rho_{AB})=\varsigma_{\min}[M],
\end{align}
where $\varsigma_{\min}[M]$ is the smallest eigenvalue of the $3\times 3$ matrix $M$ of elements
\[
M_{m,n}=\frac{1}{2}\sum_{i,l:q_i+q_l\neq0}\frac{(q_i-q_l)^2}{q_i+q_l}\bra{\psi_i}{\sigma_m}_A\otimes\mathbb I_B\proj{\psi_l}{\sigma_n}_A\otimes\mathbb I_B\ket{\psi_i}\]
with $\{q_i, \ket{\psi_i}\}$ being respectively the eigenvalues and eigenvectors of $\rho_{AB}$, $\rho_{AB} = \sum_i q_i \ket{\psi_i}\!\bra{\psi_i}$.
This provides a closed formula for discord-type correlations, quantified by the IP, of all qubit-qudit states $\rho_{AB}$. %and constitutes the only general closed formula known for a proper measure of discord, in addition to the companion one available for the LQU \cite{sLQU}.
For pure states, quantum discord is equivalent to entanglement, i.e. ${\mathcal P}^A(\ket{\psi}_{AB})$ reduces to a measure of entanglement (tangle) between $A$ and $B$.

   \subsection{Proofs that $\boldsymbol{{\cal P}^A}$ is a measure of discord-type quantum correlations}
Here we prove that the IP ${\cal P}^A$ satisfies all the established criteria to be defined as a {\it measure} of general quantum correlations \cite{smodirev}. We treat here the general case in which system $A$ is $d_A$-dimensional, while system $B$ has arbitrary (finite or infinite) dimension.
Before stating and proving the main properties of the IP, we observe that by virtue of a hierarchic inequality between the QFI and the Wigner-Yanase skew information \cite{luoskew}, it holds that \begin{equation}\label{IPvsLQU}
{\mathcal P}^A(\rho_{AB}|\Gamma) \geq 	{\mathcal U}^A(\rho_{AB}|\Gamma)
 \end{equation} for all bipartite states $\rho_{AB}$, where ${\mathcal U}^A$ denotes another recently introduced measure of discord-type correlations, known as local quantum uncertainty (LQU) and defined as \cite{sLQU}
\begin{equation}
	{\mathcal U}^A(\rho_{AB}|\Gamma) = \min_{H_A^ \Gamma} I(\rho_{AB}; H_A^\Gamma) \label{Udef}
\end{equation}
 with $I(\rho_{AB},H_A^\Gamma)=-\frac 12 \text{Tr}\Big[[\rho_{AB}^{\frac 12},H_A^\Gamma\otimes \mathbb I_B]^2\Big]$ being the skew information \cite{wigner}.
 %The inequality (\ref{IPvsLQU}) is saturated for pure quantum states

%It is easy to see that $\mathcal Q_A$ is always non negative.
\begin{Theorem}
The following properties hold:
\begin{itemize}
\item[(i)] if  $\rho_{AB}$ is a classical state (with respect to $A$) then $\mathcal P^A(\rho_{AB}|\Gamma)$ vanishes; furthermore if $\Gamma$ is nondegenerate then  $\mathcal P^A(\rho_{AB}|\Gamma)$ vanishes only on states which are classical (faithfulness criterion);
\item[(ii)] $\mathcal P^A(\rho_{AB}|\Gamma)$ is invariant under local unitary operations applied to the state $\rho_{AB}$;
\item[(iii)] $\mathcal P^A(\rho_{AB}|\Gamma)$ is monotonically decreasing under local completely positive and trace preserving maps (quantum channels) on subsystem $B$;
\item[(iv)] $\mathcal P^A(\rho_{AB}|\Gamma)$ reduces to an entanglement monotone for pure states $\rho_{AB}$.
\end{itemize}
\begin{proof} \quad \\ \vspace*{-.3cm}
\begin{itemize}
\item[(i)] If $\rho_{AB}$ is classical (with respect to $A$), $\rho_{AB}=\sum_j s_j\proj{j}_A\otimes\chi_{B,j}$, by choosing $H_A^\Gamma$ diagonal in the local basis $\ket j_A$ it is easy to see that the QFI vanishes, hence $\mathcal P^A(\rho_{AB}|\Gamma) =0$. Conversely, if $\mathcal P^A(\rho_{AB}|\Gamma) =0$ then from (\ref{IPvsLQU}) also the corresponding LQU nullifies. Under the assumption that $\Gamma$ is nondegenerate this implies that the state $\rho_{AB}$ must be classical \cite{sLQU}.
\item[(ii)] Given $\rho'_{AB} = (U_A\otimes U_B) \rho_{AB} (U_A^\dag \otimes U_B^\dag)$ with $U_A$, $U_B$ unitaries operating on $A$ and $B$ respectively, from the definition of $\mathcal{P}^A$ one has $F(\rho'_{AB};H_A^\Gamma) =F(\rho_{AB};U_A^{\dagger} H_A^\Gamma U_A)$. The invariance finally follows by noticing that the Hamiltonians  $U_A^{\dagger} H_A^\Gamma U_A$ and $H_A^\Gamma$ have the same spectrum $\Gamma$.
\item[(iii)] It follows from the properties of the QFI. Suppose $\rho'_{AB}$ is obtained from $\rho_{AB}$ by the action of a quantum channel on subsystem $B$ only. Any such a map commutes with the local phase transformation induced by $H_A^\Gamma$, hence it can be equivalently considered to be applied after the encoding stage, i.e.~to be part of the measurement process. The claim then follows by simply observing that $F(\rho_{AB};H_A^\Gamma)$ is associated with the maximum precision achievable through the optimal estimation strategy, i.e. $F(\rho_{AB};H_A^\Gamma) \geqslant F(\rho'_{AB};H_A^\Gamma)$.
\item[(iv)] The proof follows by noting that for pure states inequality (\ref{IPvsLQU}) is saturated and the QFI is proportional to the variance of the generator $H_A^\Gamma$. Then, Ref.~\cite{sLQU} shows that the minimum local variance is an entanglement monotone for pure states.
\end{itemize}
\end{proof}
\end{Theorem}
Properties (i)-(iv) imply that for all nondegenerate $\Gamma$, the quantity $\mathcal P^A$ is in fact a proper measure of quantum correlations of the discord type \cite{sOZ}.

\subsection{Examples of IP for two qubits}
Here we evaluate Eq.~(\ref{IPforS}) explicitly for a selection of two-qubit instances. Consider first the Werner states
\begin{align}
	\rho^W_{AB}=f\ket{\Phi}^{\rm Bell}\!\bra{\Phi}_{AB}+\frac{1-f}{4}\mathbb I_{AB},
\end{align}
with $0 \leq f \leq 1$.
In this case, $M = \frac{2f^2}{1+f} \mathbb I_{3\times 3}$ which implies simply ${\cal P}^A(\rho^W_{AB}) = \frac{2f^2}{1+f}$.

More generally, we can investigate two-qubit states with maximally mixed marginals, also known as Bell diagonal states, i.e.~states of the form
\begin{equation}
	\rho^M_{AB}=\frac{1}{4}\left(\mathbb I_{AB} +\!\sum_{i,j=1}^3 C_{ij}\ {\sigma_i}_A \otimes{\sigma_j}_B\right)
\end{equation}
with $C$ being the $3\times3$ real correlation matrix of elements $C_{ij}  = \tr{\rho_{AB}({\sigma_i}_A \otimes{\sigma_j}_B)}$.
Exploiting the invariance of  $\mathcal P^{A}$ under local unitaries to express $C$  in terms of its singular values $\{c_1, c_2, c_3\}$, we find that
\begin{equation}\label{IPBD}
\mathcal P^{A}(\rho_{AB}^M)=  \frac{\|C\|_2^2 -\|C\|^2_{\infty}+ 2\,\det C}{1-\|C\|^2_{\infty}}\,,
\end{equation}
where $\| C \|_2^2 = \tr{C^T C} = c_1^2+c_2^2+c_3^2$ and $\|C\|_{\infty}^2 = \max\{c_1^2,c_2^2,c_3^2\}$ are, respectively, the squared Hilbert-Schmidt and operator norms of $C$. One can verify that the IP behaves similarly to other measures of discord under typical dynamical conditions \cite{smodirev}, e.g., it exhibits freezing under nondissipative decoherence in the specific settings which have been identified as universal for all valid measures of discord \cite{ben}.

For two-qubit separable states, we find that the IP can reach as high as $\frac12$, as it is the case, for instance, for the state $\rho_{AB}^{\rm sep}=(\ket{0}_A\ket{0}_B\bra{0}_A\bra{0}_B + \ket{{+}}_A\ket{1}_B\bra{{+}}_A\bra{1}_B)/2$.

\section{Experimental NMR implementation}

\subsection{NMR setup}

%\subsection{Basics of NMR quantum information processing}
In this work, we performed the experiments in a room temperature NMR setup
using a liquid-state sample of $^{13}$C-labeled chloroform molecules (CHCl$_{3}$). %, containing 9.99\% of the $^{13}$C isotope.
In such a sample, the intramolecular
interaction is given by the scalar $J$ coupling  \cite{livro_abragam, livro_ernst,livro_ivan}, which is a Fermi contact interaction
type, due to the superposition of the electron wave functions of the carbon and hydrogen
atoms in the CHCl$_3$ molecule. The (intramolecular and intermolecular) dipolar
interaction between the spins is instead cancelled in the average, due to the random orientations
of the molecules in the liquid sample.
%On the other hand, the direct
%dipolar interaction between different molecules is
%negligible compared to the intramolecular interactions, due to the fact that we are dealing with
%a solution with small CHCL$_3$ concentration.
This implies that one can effectively consider the
sample as an ensemble of independent molecules \cite{rmp}.
A two-qubit system is therefore constituted by the ensemble of the two nuclear spins of the  $^{1}$H (qubit $A$) and $^{13}$C (qubit $B$) atoms in each molecule  (see Fig.~\ref{figs2}{\abl{a}}),
 coupled through the scalar $J$  interaction \cite{livro_abragam, livro_ernst,livro_ivan}.

In the rotating frame at the resonant frequency for each species of nuclei ($\omega^{\rm H}/2\pi \approx $ 500 MHz for $^{1}$H and
$\omega^{\rm C}/2\pi \approx $ 125 MHz  for $^{13}$C), the nuclear spin Hamiltonian is given by
\begin{equation}\label{hamtot}
\mathcal{H} =
%-\delta^{\rm H}\,I_{z}-\delta^{\rm C}\,S_{z}+
2\pi \,J\,I_{z}\,S_{z} +\omega _{1}^{\rm H}\,(I_{x}\cos \phi ^{\rm H}+I_{y}\sin \phi ^{\rm H})+\omega
_{1}^{\rm C}(S_{x}\cos \phi ^{\rm C}+S_{y}\sin \phi ^{\rm C})\,,
\end{equation}%
where $I_{\alpha }$, $S_{\beta }$ are the spin angular momentum operators in
the $\alpha ,\beta =x,y,z$ direction for the $^{1}$H and the $^{13}$C nuclei, respectively,
%(equal to half of the corresponding Pauli matrices ${\sigma_\alpha}_A, {\sigma_\beta}_B$),
$\phi^{\rm H}$ and $\phi^{\rm C}$ define the direction of the radiofrequency (rf)  field (pulse phase),
and $\omega _{1}^{\rm H}$ and $\omega _{1}^{\rm C}$ are the rf nutation
frequency (rf power) for the nuclei.
%The first two terms in Eq.~(\ref{hamtot})
%describe the chemical shifts for the $^{1}$H and $^{13}$C nuclei, respectively.
%The resonance frequencies are $\omega^{\rm H}/2\pi \approx $ 500 MHz and
%$\omega^{\rm C}/2\pi \approx $ 125 MHz.
The first term in Eq.~(\ref{hamtot}) is due to a scalar spin-spin
coupling of $J\approx $ 215 Hz. The second and third terms represent the rf
field to be applied to the $^{1}$H and $^{13}$C nuclear spins, respectively.

At room temperature $T$, the ratio between magnetic and thermal energies of each qubit is $\epsilon =\frac{\hbar \omega _{L}}{4\,k_{B}\,T}\sim 10^{-5}$, where $\omega_L$ is the Larmor frequency.
The Gibbs state can therefore be expanded as
\begin{equation}
\rho \approx \frac{\mathbb{I}}{4}+\epsilon \Delta \rho ,
\end{equation}%
where  the traceless term $\Delta \rho $, called deviation density
matrix, contains all the information about the state of the system.
Every unitary manipulation of the qubits --- done using rf pulses,
evolution under the coupling interaction and magnetic field gradients --- affects only the deviation matrix,  leaving the identity unaffected. For the needs of quantum information processing protocols \cite{livro_ivan},
 it is customary to introduce a normalization of the NMR data so as to obtain so-called pseudo-pure states \cite{1997_Science_275_350,1997_pnas_94_1634}, which in the case of a two-qubit system take the form
\begin{equation}\label{pseudo}
\rho =\frac{(1-\epsilon )}{4}\mathbb{I}+\epsilon \rho_{AB}.
\end{equation}%
where $\rho_{AB}$ is a density operator, related to the deviation matrix by $%
\rho_{AB}=\Delta \rho +\mathbb{I}/4$. By  state preparation in NMR, we refer henceforth to engineering a desired target state  in the density operator $\rho_{AB}$. Although arbitrary states, including pure and entangled ones, can be encoded in $\rho_{AB}$ in this way, recall that the physical density matrix $\rho$ is always separable because its deviation from the totally mixed state is as small as $\epsilon \sim 10^{-5}$ \cite{1999_PRL_83_1054,2001_PRL_87_047901}.
%, and the physical quantities calculated using the pseudo-pure description are always given in order of $\epsilon$.

To realize the experiment, we employed hard radiofrequency pulses, transverse to the
static magnetic field, on the resonance frequencies of $^{1}H$ and $^{13}C$ nuclear spins. Furthermore,
the two-qubit gates use free evolution periods under the interaction Hamiltonian to
be implemented. They are designed based on the controlled phase gate, given by the
sequence $U_{J}\left(\frac{1}{2J}\right){\sf Z}_{A}{\sf Z}_{B}$, where $U_{J}$ is the free evolution
operator under the interaction Hamiltonian and the ${\sf Z}_i$'s are $\left(\frac{\pi}{2}\right)$
rotations around the $z$ axis for the $i$-th qubit. The pulse sequences are optimized using commutation relations
between the rotations, in order to remove all the $z$ rotations from steps of the sequence
where the density matrix is diagonal. Since diagonal states are invariant under $z$
rotations, the $z$ rotations do not need to be implemented.

The detection in NMR is done independently for each nuclear species by measuring the free induction decay of the response signal of the sample,
that is, by reading the transverse magnetization, after
the excitation of the sample via rf pulses, given e.g.~for qubit $A$ by $M_{\bot
}(t)=M_{x}(t)+iM_{y}(t)$, where $M_{\alpha }=\langle I_{\alpha }\rangle =%
\tr{[(I_{\alpha } \otimes \mathbb{I})\,\rho]}=\epsilon \,\text{Tr}[(I_{\alpha } \otimes \mathbb{I})\,\rho _{AB}]$ (%
$\alpha =x,y$) \cite{livro_abragam, livro_ernst}. Note
that the magnetization is proportional to $\epsilon$ and that the
measurements are over the whole molecular ensemble \cite{livro_ivan,snmrmetro}.
The Fourier transform of the
magnetization signal gives the NMR spectra of the nucleus. For the system
considered in this work, describing the density operator in the product
operator basis \cite{Jones_review}, the readout of a measurement, say on qubit $A$, can be
expressed  by the matrix relation
\begin{equation}
\left(
\begin{array}{c}
\text{Tr}[\rho \tilde{I}_{+}\otimes \mathbb{I}] \\
\text{Tr}[\rho \tilde{I}_{+}\otimes \tilde{S}_z]%
\end{array}%
\right) =\left(
\begin{array}{cc}
1 & 1 \\
1 & -1%
\end{array}%
\right) \left(
\begin{array}{c}
S(\omega _{1}-\pi J) \\
S(\omega _{1}+\pi J)%
\end{array}%
\right) ,  \label{09b}
\end{equation}
\noindent where $S(\omega_1 \pm\pi J)$ are the intensities of the NMR
spectrum at the frequencies given by $\omega_1 -\pi J$ and $\omega_1 +\pi J$.
The measurement operator is given by $\tilde{I}_{+}$ $=$ $\tilde{I}_x +i\tilde{I}_y$,
and $\tilde{I}_{\alpha}$ $=$ $R^{-1}I_{\alpha}R$, where $R$ is a $\pi/2$ reading pulse.
%that projects the magnetisation on the $xy$ plane.
The set of reading pulses is given by
Pauli gates \cite{nielsenchuang} $\sf X$, $\sf Y$ and $\sf I$, which are rotations in the $x$ or $y$ directions, or respectively no pulse. Here the density matrix $\rho$ denotes the system state before the reading pulse.

By applying a particular set of reading pulses (in our case, the pulses $\sf II$, $\sf IX$, $\sf IY$, $\sf XX$), we can perform quantum state tomography of the
full density operator of the two-spin system \cite{1998_PRSA_454_447,2004PRA}, thus reconstructing the produced state $\rho_{AB}$. From the tomographically reconstructed density matrices, we can calculate all the physical quantities of interest.

%--------------------------------------------------------------------------------

Small pulse imperfections due to the electronic devices and field inhomogeneity induce an experimental error of  less than 0.3\% per pulse.
%The total uncertainty of a particular NMR experiment must include the error in the preparation of the initial states, in the
%implemented protocols and in the quantum state tomography process, besides errors caused by relaxation.
For our experimental setup, we estimate a global error of less than 5\%.

\subsection{Probe state preparation}

\begin{figure}[t]
\begin{center}
\includegraphics[width=3.5cm]{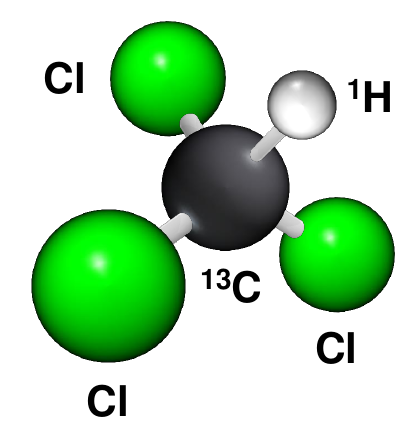}
\end{center}
\caption{\label{figaclora}
Graphical representation of the chloroform molecule. The $^1$H nuclear spin-$\frac12$ realizes qubit $A$ and the $^{13}$C nuclear spin-$\frac12$ realizes qubit $B$.}
\end{figure}

The sample used in the experiments was a solution of 100 mg of 99.99\% $^{13}$C-labelled CHCl$_{3}$ (see Fig.~\ref{figaclora})
dissolved in 0.7 ml of 99.8\% $d_6$ acetone, both compounds provided by the Cambridge Isotope Laboratories Inc.,
 inserted in a 5 mm NMR tube. Taking into account the molar mass of chloroform ($119.38$ g/mol) and the factor $\epsilon \sim 10^{-5}$, we can estimate a number $\nu \approx 10^{15}$ of independent polarized molecules in our ensemble.
The experiments were carried out at room temperature in a
Varian 500 MHz Premium Shielded spectrometer, located at the Brazilian Center for Research in Physics (CBPF, Rio de Janeiro)
using a Varian 5 mm
double resonance probehead equipped with a magnetic field gradient coil. The
spin lattice relaxation times, measured by the inversion recovery pulse
sequence, were $3.57$s and $10$s for $^{1}$H and $^{13}$C, respectively. The
spin-spin relaxation times, measured by a CPMG pulse sequence, were $1.2$s and  $0.19$s for $^{1}$H and $^{13}$C, respectively. Appropriate pulse sequences were applied to engineer the states $\rho^C_{AB}$ and $\rho^Q_{AB}$ as described in Fig.~\ref{figs2}.

%For the black box implementation, in all our experimental runs, we set the actual value of the phase $\varphi$  at  $\varphi_0=\frac{\pi}{4}$. The following %Table summarises the chosen Hamiltonians $H_A^{(k)}$ and the pulse sequences we implemented to perform the corresponding rotations $U_A^{(k)}$.

\subsection{State tomography}

\begin{figure}[t]
\begin{center}
\includegraphics[width=.8\textwidth]{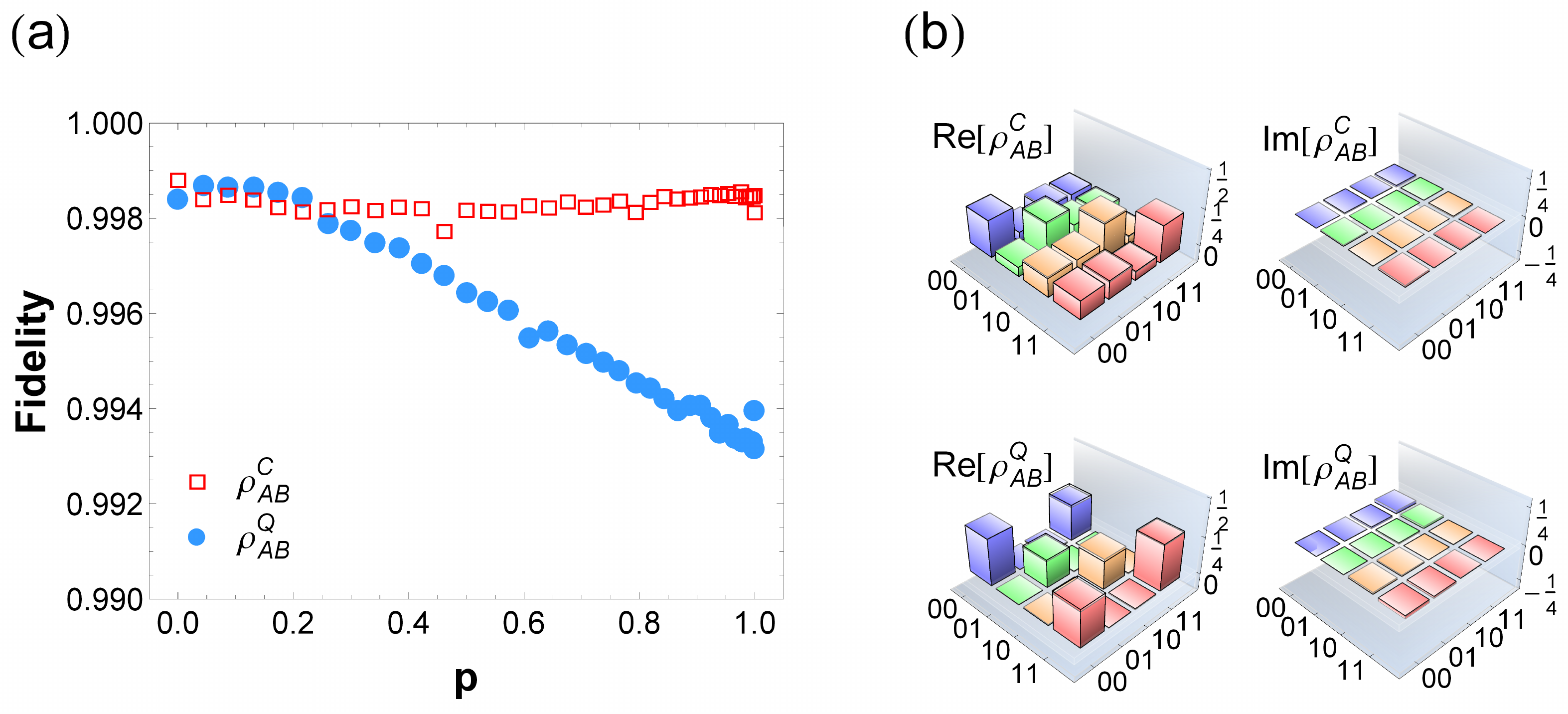}
\end{center}\caption{Quantum state tomography of the input probe states $\rho_{AB}^{C,Q}$. Panel \abl{a} shows the fidelities with the theoretical states defined in the main text, as calculated from the formula (\ref{fidelity}), for all the acquired values of $p$. The obtained fidelities are all well above $0.99$. Panel \abl{b} shows an instance of tomographically reconstructed density matrices for classical and discordant probes with $p=0.5$. The nearly indistinguishable darker bar edges correspond to the theoretical predictions.\label{figs3}}
\end{figure}

We performed full state tomography in order to reconstruct the prepared states $\rho_{AB}^{C,Q}$ and compared them with the theoretical prescriptions. We obtained an excellent agreement (see Fig.~\ref{figs3}) quantified by a  state fidelity $\cal F$, averaged over all the different settings for $p$ across both families of states, as high as $(99.7 \pm 0.2)\%$. Here we used the Hilbert-Schmidt fidelity \cite{hfidelity} conventionally adopted in NMR experiments and defined by
\begin{equation}\label{fidelity}
{\cal F}(\rho_{AB}^{\rm th},\rho_{AB}^{\rm exp}) = \frac{\tr{\rho_{AB}^{\rm th}\ \rho_{AB}^{\rm exp}}}{\sqrt{\tr{(\rho_{AB}^{\rm th})^2}\ \tr{(\rho_{AB}^{\rm exp})^2}}}\,,
\end{equation}
where  $\rho_{AB}^{\rm th}$ and $\rho_{AB}^{\rm exp}$ denote respectively the theoretical and experimental density matrices for each given setting.

\subsection{Choice of the black box settings}
Given the iso-purity two-qubit states $\rho_{AB}^C$ and $\rho_{AB}^Q$ defined in the main text, one can analyze theoretically which black box settings would be the most favourable or, respectively, the most adverse for the estimation of a parameter $\varphi$ encoded in a unitary shift on qubit $A$ of the form $U_A=e^{-i \varphi H_A}$. We have studied the QFI of the chosen probe states as a function of the spherical coordinate angles $(\vartheta,\phi)$ which define a generic single-qubit nondegenerate Hamiltonian $H_A = \vec{n}_A \cdot \vec{\sigma}_A$, with $\vec{n}_A=(\sin\vartheta\cos\phi, \sin\vartheta\sin\phi,\cos\vartheta)$, see a detail in Fig.~\ref{figsfes}. It is found for any value of $p\in (0,1)$ that, modulo periodicities, both QFIs $F(\rho^{C,Q}_{AB}; H_A)$ are maximized at $\vartheta=0$; furthermore,  $F(\rho^{Q}_{AB}; H_A)$ is minimized at $\vartheta=\pi/2$ and any value of $\phi$, while  $F(\rho^{C}_{AB};H_{A})$ is minimized at $\vartheta=\pi/2$ and $\phi=0$. Consequently, the chosen settings include the best case scenario for both probes (setting $k=1$, corresponding to $\vartheta=0$), the worst-case scenario for both probes (setting $k=3$, corresponding to $\vartheta=\pi/2$ and $\phi=0$) and an intermediate case (setting $k=2$, corresponding to $\vartheta=\pi/2$ and $\phi=\pi/4$, which is again a  worst-case scenario for the discordant probes but not for the classical ones.

\begin{figure}[h!]
\begin{center}
\includegraphics[width=8.5cm]{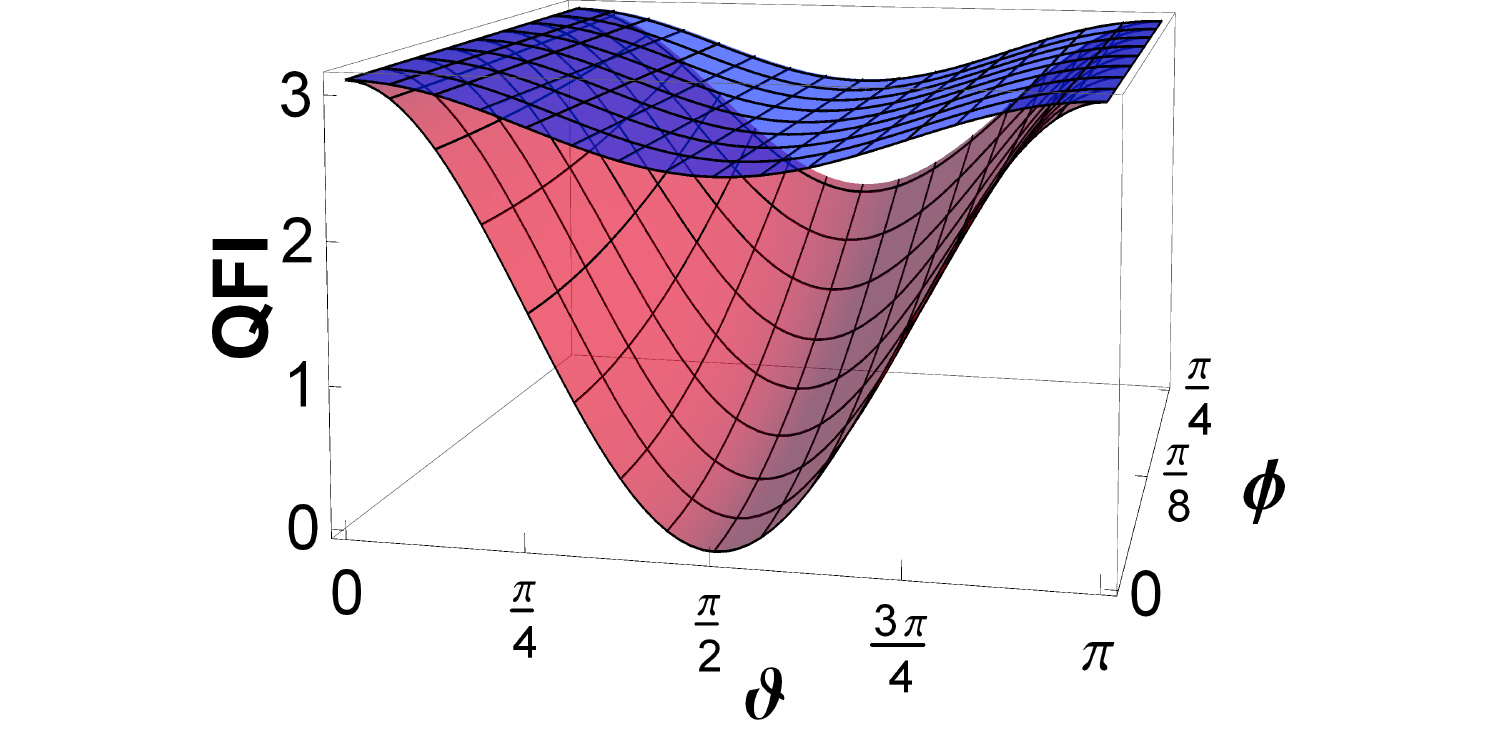}
\end{center}
\caption{\label{figsfes}
QFIs for discordant probes $\rho_{AB}^Q$ (blue) and for classical probes $\rho_{AB}^C$ (red), with $p=0.8$, as a function of the local Hamiltonian angular parameters $\vartheta,\phi$. The chosen black box settings as implemented in the experiment are detailed in the text.
}
\end{figure}

\subsection{Optimal measurement and readout}
Following the implementation of a black box unitary shift,  each particular input state  $\rho_{AB}$ is transformed into a (generally) $\varphi$-dependent state $\rho_{AB}^{\varphi}$. The optimal detection strategy to infer the value of $\varphi$ with the maximum allowed precision, once the chosen setting $k$ is disclosed, consists in measuring $\rho^\varphi_{AB}$ in the eigenbasis $\{\ket{\lambda^\varphi_j}\}$ of the symmetric logarithmic derivative (SLD) operator \cite{squantumcramer}
\begin{equation}\label{SLD}
L^{(k)}_{\varphi}(\rho^\varphi_{AB}) = \sum_j l_j \ket{\lambda^\varphi_j}\bra{\lambda^\varphi_j}=2\sum_{i,j:q_i+q_j \neq 0}\frac{\bra{\psi_i}\partial_{\varphi}\rho^{\varphi}_{AB}\ket{\psi_j}}{q_i+q_j}\ket{\psi_i}\bra{\psi_j}.
\end{equation}
In general, the eigenvectors of the SLD depend on the initial probe state and on $H_A^{(k)}$ but also on the value of the unknown parameter $\varphi$. To be able to saturate the quantum Cram\'er-Rao bound asymptotically, one possible metrological strategy is to resort to an adaptive scheme, where a fraction of the available probes are iteratively consumed to run a rough estimation localizing the true value of $\varphi$, which is then fedforward to adjust the measurement basis on the remaining probes \cite{sparis}. In our NMR implementation, we have a very large number $\nu \approx 10^{15}$ of independent probes (each being one polarized molecule in the sample) but it is not possible to address them individually. Every measurement amounts to a spatial average of the whole sample, which by ergodicity is equivalent to a temporal average of $\nu$ consecutive measurements in a single run \cite{snmrmetro}. In principle, one could repeat the experiment with the whole ensemble (and for all possible settings) several times, adjusting the measurement basis after each implementation and seeking for convergence towards the eigenbasis of the SLD. However, we adopted a less demanding procedure described as follows.

Based on the experimentally reconstructed input probe states $\rho_{AB}^{C,Q}$, we ran a numerical simulation of an iterative adaptive procedure, to localize the value of $\varphi_{\rm trial}$ which could then be used for the  selection of the SLD $L^{(k)}_{\varphi_{\rm trial}}$ in the remainder of the experiment. We observed a rapid convergence to $\varphi_{\rm trial} \approx \varphi_0=\frac{\pi}{4}$ in all nontrivial settings (see Fig.~\ref{figs4}). We set therefore  $L^{(k)}_{\varphi_0}$ as our design SLD operator defining the optimal measurement strategy to be implemented experimentally for all settings. Notice that in the particular case of $\rho_{AB}^C$ for $k=3$ the SLD is independent of $\varphi$ so this analysis is not necessary.

\begin{figure}[t]
\begin{center}
\includegraphics[width=.43\textwidth]{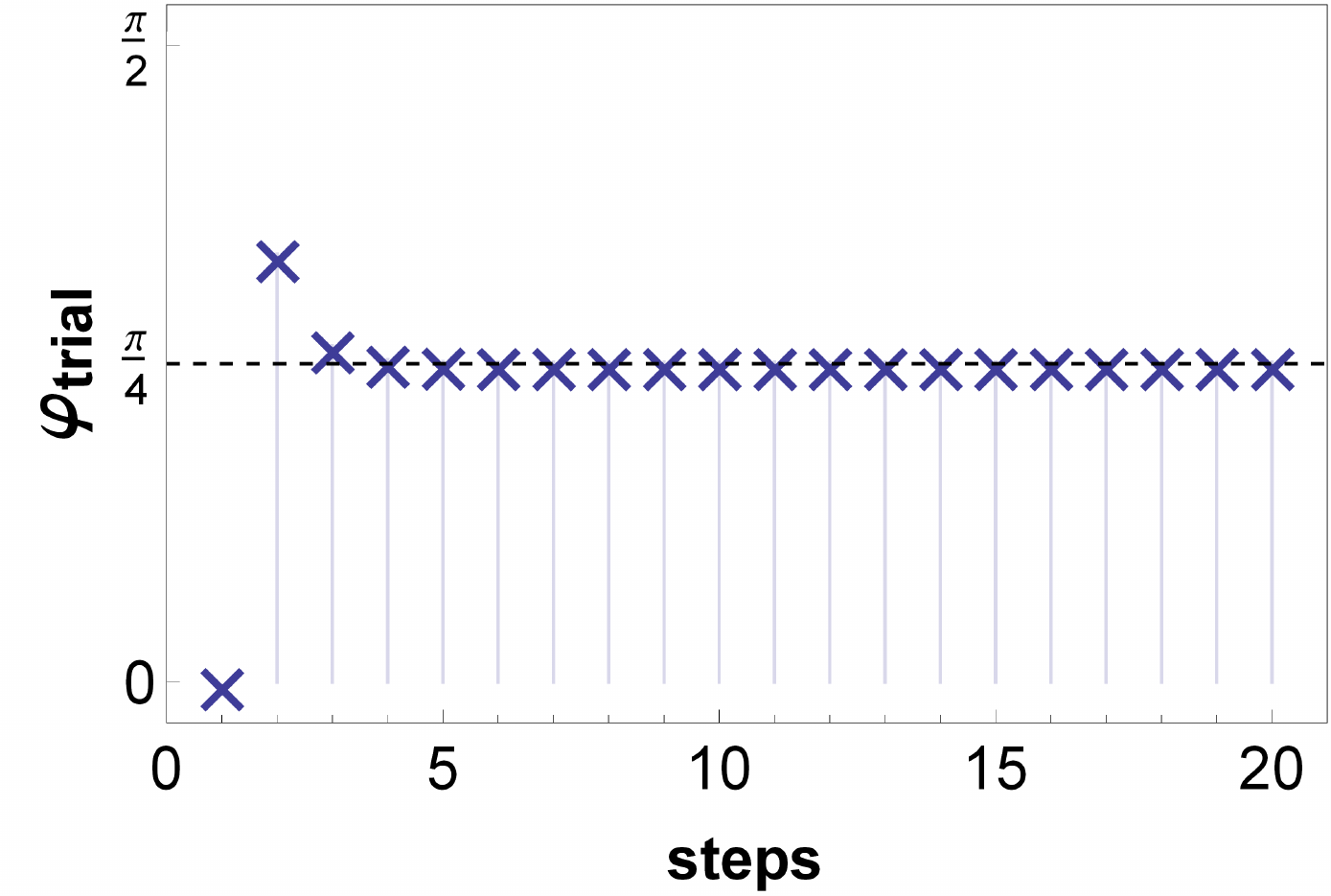}
\end{center}\caption{Snapshot from the simulation of adaptive phase estimation, plotted for the instance of an input probe $\rho_{AB}^Q$, with $p=0.13$, and for the black box setting $k=1$. The experimentally reconstructed density matrix for the input probe state is used as a basis for the simulation. At each step $n$, we simulate an ensemble measurement in the eigenbasis of the SLD $L^{(k)}_{\varphi_{\rm trial}(n)}$ following a rotation of the probe state by $\varphi_0=\frac{\pi}{4}$ as generated by $H_A^{(k)}$.   To initialize the procedure we choose without loss of generality $\varphi_{\rm trial}(1)=0$. The output data from the simulated ensemble measurement are processed by a least-squares method, analogous to that used to analyze the actual experimental data in the main text, in order to infer the expectation value of the estimator $\langle \tilde{\varphi}_{\rm trial}(n)\rangle$, which is then used to setup the next step, $\varphi_{\rm trial}(n+1)\equiv\langle \tilde{\varphi}_{\rm trial}(n)\rangle$. We observe a very rapid convergence of $\varphi_{\rm trial}$ towards the true value $\varphi_0$, which is therefore adopted to set the actual measurement procedure in the experiment.
%Similar results are found for other values of $p$ in the input states and different settings $k$ (apart from the trivial case $k=3$ for classical probes).
}\label{figs4}
\end{figure}

To implement the projections on the SLD eigenstates, as shown in Fig.~\ref{figs2} (green panels), we first applied a global change of basis transformation, described by a matrix $V_{k}$ (for each setting $k$) whose rows are the eigenbras $\{\bra{\lambda^{\varphi_0}_j}\}$ of the corresponding SLD $L^{(k)}_{\varphi_0}$. This transformation, which is different for classical and discordant probes, served the purpose to map the SLD eigenvectors onto the computational basis of the two qubits, so that the ensemble expectation values $d_j^{\rm exp} = \bra{\lambda_j^{\varphi_0}} \rho_{AB}^\varphi \ket{\lambda_j^{\varphi_0}}$ could be directly observed in the diagonal elements of the output density matrix.
The application of a pulsed field gradient ${\sf G}_z$ completes the optimal detection. The populations of the density matrix were finally measured by an appropriate set of reading pulses.

The complete set of output measured data, corresponding to the ensemble expectation values $d_j^{\rm exp}$ for each setting $(k,s)$, is reported in Fig.~\ref{figs5}.

The Table~\ref{tabs2}  on the final page collects,
 for each combination $(k,s)$ of input probe state $\rho_{AB}^{s}$ ($s=C,Q$) and black box setting ($k=1,2,3$), the  eigenvalues $l_j$ of the respective SLD (which are independent of $\varphi$) and the corresponding eigenbasis $\{\bra{\lambda^{\varphi_0}_j}\}$ as encoded in the matrix $V_k^{s}$; the quantum circuits for the implementation of $V_k^{s}$ via standard gates (Pauli gates $\sf X$, $\sf Y$, $\sf Z$, Hadamard gate $\sf H$, phase gates ${\sf P}_\phi = \text{diag}(1,e^{i \phi})$, and {\sf CNOT}) \cite{nielsenchuang} are shown as well.

\begin{figure}[t]
\begin{center}
\includegraphics[width=.75\textwidth]{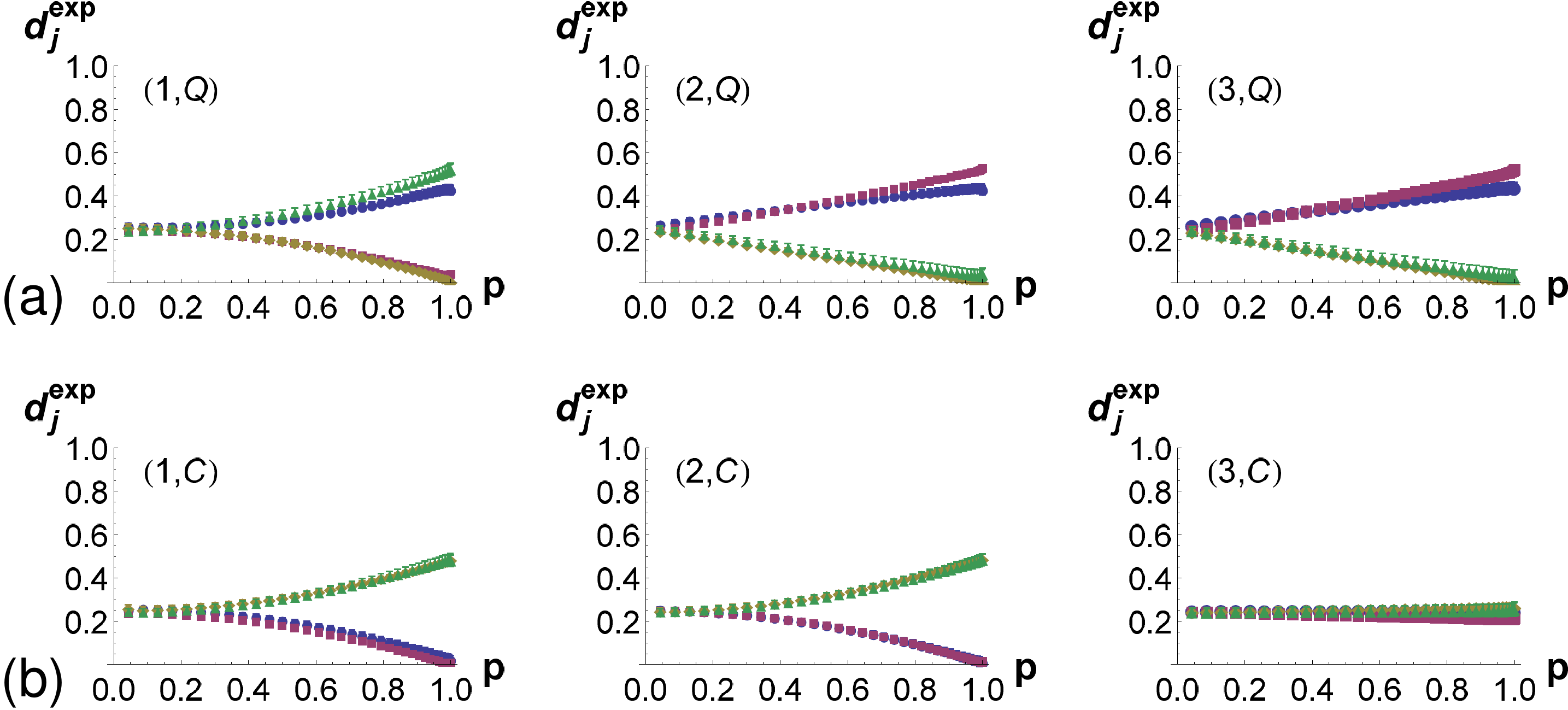}
\end{center}\caption{Experimental data. Measured expectation values $\{d^{\rm exp}_j\}$ of the output states in the eigenbasis of the  SLD for all the settings $(k,s)$ encompassed in our demonstration. Each row refers to one of the two families of (discordant vs classical) probes, with $s=Q,C$ for rows \abl{a}, \abl{b}, respectively; on the other hand, each column represents a different black box setting $k=1,2,3$ implemented in the protocol. Referring to the ordering of the SLD eigenvectors as given in Table~\ref{tabs2}, the plot legend in all panels is: \textcolor[rgb]{0.2472, 0.24, 0.6}{\ding{108}} $d_1^{\rm exp}$, \textcolor[rgb]{0.6, 0.24, 0.442893}{\ding{110}} $d_2^{\rm exp}$, \textcolor[rgb]{0.6, 0.547014, 0.24}{\ding{117}} $d_3^{\rm exp}$, \textcolor[rgb]{0.24, 0.6, 0.33692}{\ding{115}} $d_4^{\rm exp}$. Error bars amounting to $5\%$ on the measured data as estimated from the pulse imperfections are included.\label{figs5}}
\end{figure}

%\clearpage

\begin{table}[th!]
\begin{center}
\includegraphics[scale=1]{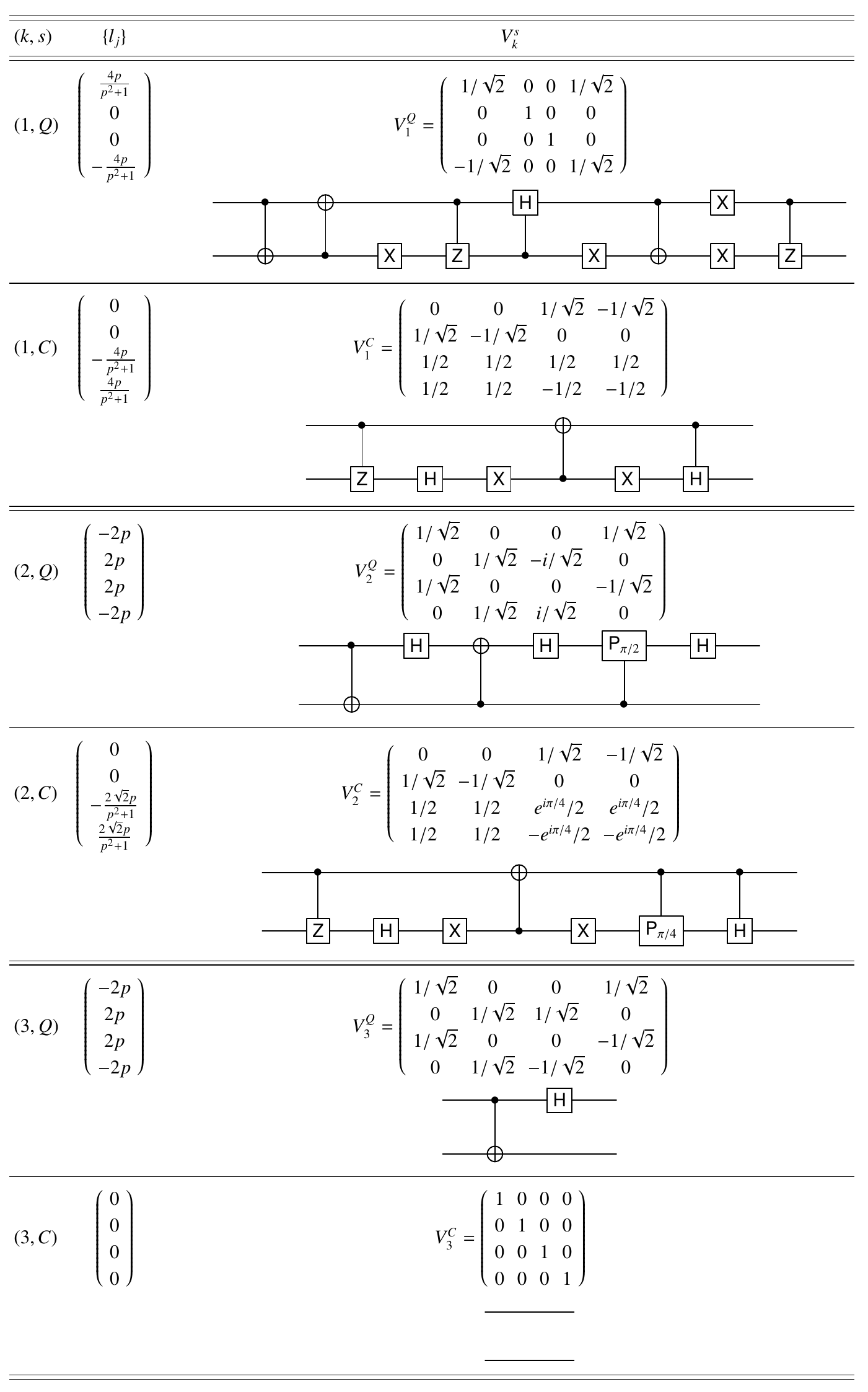}
\caption{\label{tabs2} Specifics of the SLD for the optimal measurement in the final step of the metrology experiment.}
\end{center}
\end{table}

%\clearpage

%\subsection{Calculation of the estimator from the measured data}

%\vfill

\clearpage
\end{widetext}


\begin{thebibliography}{99}
\bibitem{nmrnatural} This setting is very natural for implementations such as NMR, where measurements on an ensemble of $\nu$ probes are taken collectively, rather than iteratively \cite{morton,nmrmetro,cory}.
\bibitem{notaav}
Here and in the following the averages are intended to be calculated on the transformed  states $\rho_{AB}^{\varphi\,s}$ after the black box.
\end{thebibliography}
\end{document}